\documentclass[aps,nofootinbib,prd,eqsecnum,showpacs,showkeys,preprintnumbers]{revtex4-1}
\usepackage{textcomp}
\usepackage{amsmath}
\usepackage{amssymb}

\makeatletter

\usepackage{graphicx}
\usepackage{amsfonts}
\usepackage{color}
\usepackage{bm}
\usepackage{mathrsfs}
\usepackage{epstopdf}
\usepackage{url}
\usepackage{footnote}
\usepackage{ulem}

\newcommand*{\rom}[1]{\expandafter\@slowromancap\romannumeral #1@}

\newcommand{\be}{\begin{equation}}
  \newcommand{\ee}{\end{equation}}
\newcommand{\ben}{\begin{eqnarray*}}
  \newcommand{\een}{\end{eqnarray*}}
\newcommand{\bea}{\begin{eqnarray}}
  \newcommand{\eea}{\end{eqnarray}}
\newcommand{\bdm}{\begin{displaymath}}
  \newcommand{\edm}{\end{displaymath}}
\newcommand{\ba}{\begin{align}}
  \newcommand{\ea}{\end{align}}

\makeatother

\begin{document}

\title{The quantum realm of the ``Little Sibling'' of the Big Rip singularity}

\author{Imanol Albarran $^{1,2}$}
\email{imanol@ubi.pt}

\author{Mariam Bouhmadi-L\'{o}pez $^{1,2,3,4}$}
\email{{\mbox{mbl@ubi.pt (On leave of absence from UPV and IKERBASQUE.)}}}

\author{Francisco Cabral $^{1,2}$}
\email{ftoc@ubi.pt}

\author{Prado Mart\'{\i}n-Moruno$^{5}$}
\email{pradomm@ucm.es}

\date{\today}

\affiliation{
${}^1$Departamento de F\'{\i}sica, Universidade da Beira Interior, 6200 Covilh\~a, Portugal\\
${}^2$Centro de Matem\'atica e Aplica\c{c}\~oes da Universidade da Beira Interior (CMA-UBI), 6200 Covilh\~a, Portugal\\
${}^3$Department of Theoretical Physics, University of the Basque Country UPV/EHU, P.O. Box 644, 48080 Bilbao, Spain\\
${}^4$IKERBASQUE, Basque Foundation for Science, 48011, Bilbao, Spain\\
${}^5$Departamento de F\'{\i}sica Te\'orica I, Ciudad Universitaria, Universidad Complutense de Madrid, E-28040 Madrid, Spain
}

\begin{abstract}

We analyse the quantum behaviour of the ``Little Sibling'' of the Big Rip singularity (LSBR) \cite{Bouhmadi-Lopez:2014cca}. The quantisation is carried within the geometrodynamical approach given by the Wheeler--DeWitt (WDW) equation. The classical model is based on a Friedmann--Lema\^{i}tre--Robertson--Walker Universe filled by a perfect fluid that can be mapped to a scalar field with phantom character. We analyse the WDW equation in two setups. In the first step, we consider the scale factor as the single degree of freedom, which from a classical perspective parametrises both the geometry and the matter content given by the perfect fluid. We then solve  the WDW equation within a WKB approximation, for two factor ordering choices. On the second approach, we consider the WDW equation with two degrees of freedom: the scale factor and a scalar field. We solve the WDW equation, with the Laplace--Beltrami factor-ordering, using a Born--Oppenheimer approximation. In both approaches, we impose the DeWitt (DW) condition as a potential criterion for singularity avoidance. We conclude that in all the cases analysed the DW condition can be verified, which might be an indication that the LSBR can be avoided or smoothed in the quantum approach. 

\end{abstract}


\maketitle

\section{Introduction}

From the early thinkers that devoted their reasoning to the nature
of motion and the related conceptions of space and time, to modern
theories of gravitation, geometry has always been an essential aspect
of the epistemological bridge between observations and theory. General
relativity (GR) opened the door for a dynamical role of space-time
geometry in physics.

GR was the main theoretical foundation for the development of relativistic
astrophysics and modern cosmology. Accordingly, the evolution of black
hole research inspired the formulation of the singularity theorems
of Hawking and Penrose \cite{Penrose} (together with the cosmic censorship conjecture,
excluding the possibility of naked singularities \cite{Penrose and Hawk}). These efforts contemplated past and future singularities. The first, associated
with the Big Bang, were supposed
to obey the Weyl curvature hypothesis, following thermodynamical motivations \cite{Penrose and Hawk},
whereas future singularities in that time were of two types: astrophysical
(gravitational collapse of a star into a black hole) and cosmological
(big crunch scenarios). On the other hand, the remarkable evolution
of research in cosmology, both theoretically and observationally, led
to the most accepted paradigm of ($\Lambda$CDM) cosmology (with inflation),
which still contains the unresolved puzzles of dark matter (DM)
and dark energy (DE) \cite{Sahnidmde,Weinberg:2000yb,Peebles:2002gy,Padmanabhan:2002ji,Copeland:2006wr,Frieman:2008sn}. The empirical data supporting a present accelerated expansion (see for example \cite{Ade:2015xua}), motivated
the re-appearance of the cosmological constant \cite{Weinberg:2000yb,Peebles:2002gy,Padmanabhan:2002ji,Sahni:1999gb}. Nevertheless, the
absence for a corresponding physical theory (or an inconsistency with
the quantum field predictions for the vacuum energy) remains. Moreover, a constant dark energy density gives the
so called coincidence problem, i.e. the same order of magnitude at present time for the DM and DE densities. These issues inspired alternative descriptions. Dynamical dark energy models based on scalar fields were widely investigated and the research is still active.

From the point of view of physical ontology, the ideal study would follow from a fundamental theory describing
the form of the equation of state for a DE fluid, based on the understanding of its origin and constituents. In fact, different cosmological evolutions can be explored depending on the nature of the energy-momentum tensor and on the equation of state (satisfying or transgressing some energy conditions). Until the present date, the nature of dark energy remains unclear, therefore cosmologists have been somewhat limited to testing different equations of state (and different potentials for the hypothetical scalar fields), and compute the resulting cosmic dynamics. 
In this context, dynamical DE models based on scalar fields have many applications. Some models try to unify DM and DE as in the case of the generalised Chaplygin gas \cite{Kamenshchik:2001cp,Bento:2003dj,Bouhmadi-Lopez:2004me} and also early-time and late-time acceleration, see for example \cite{Nojiri:2005pu}. Other models assume an interaction between the DM and DE components (see \cite{Costa:2013sva} and \cite{Salvatelli:2014zta} for recent observational constraints). But most importantly, was the discovery that dynamical dark energy models known as phantom energy (which might have a negative kinetic term) could give rise to a new type of singularity, the Big Rip \cite{StarobBigrip,Caldwellbigrip,Caldwellbigrip2,Caldwell:2003vq,Carrollbigrip,Chimentobigrip,Dabrbigrip,Gonzalezbigrip,Gonzalezbigrip2}, in which the scale factor, the Hubble parameter and its cosmic time derivative diverge in a finite cosmic time. It was the beginning of the study of late-time or future singularities, or more generally, of cosmic singularities related to dark energy. Other types of singularities were discovered such as, Sudden singularity \cite{Barrowbigbrake,Gorinibigbrake,Nojirisudden,Barrowsudden}, Big Freeze  \cite{Nojirisudden,Nojiribigfreeze,Nojiribigfreeze2,Mariambigfreeze,Mariambigfreeze2},
type IV \cite{Nojiribigfreeze2,Nojiribigfreeze,Nojirisudden,Mariambigfreeze,Mariambigfreeze2,Nojiritypeiv,Bambatypeiv,Nojiri:2015fia}, Little Rip \cite{Ruzmaikina1970,Stefancic:2004kb,BouhmadiLopez:2005gk,Bouhmadi-Lopez:2013nma,Frampton:2011sp,Brevik:2011mm} and the recently investigated Little Sibling of the Big Rip (LSBR) \cite{Bouhmadi-Lopez:2014cca}.

Future cosmic singularities or abrupt events are fascinating new areas of cosmology. Can these events be absent when the classical theory is quantised? Some works have applied the methods of quantum cosmology to the classical models leading to these DE related singularities \cite{Bouhmadi-Lopez:2009pu,Bouhmadi-Lopez:2013tua,Albarran:2015tga,Kamenshchik:2007zj,Dabrowski:2006dd,deHaro:2012xj,Barrow:2011ub,Nojiri:2015fia,Barrow:2015sga,Elizalde:2004mq,Bamba:2012ka}. In fact, dynamical dark energy models based on scalar fields (quintessence or phantom) provide a very interesting and suitable scenario to explore the quantum gravity challenge. Accordingly, in quantum cosmology the restriction to an isotropic and homogeneous universe, simplifies substantially the general theory. The approach is done in the line of quantum geometrodynamics where a canonical quantisation of gravity (metric functions and conjugate momenta) is performed \cite{KieferQG,PMonizQC}. The main purpose is to solve the Wheeler--DeWitt (WDW) equation and apply appropriate boundary conditions, in order to get the wave function of the Universe \cite{KieferQG,PMonizQC,FuturePhscs}. In \cite{Dabrowski:2006dd} it was shown that after solving the WDW equation with phantom dark energy with a corresponding exponential potential, the quantum effects dominate the region of the classical Big Rip singularity, where classically the scale factor, the Hubble rate and it cosmic time derivative blow up at a finite future cosmic time. These authors found wave packets solutions that follow the classical trajectory and disperse in the genuinely quantum region. These quantum effects occur at large scales and since the solutions are regular the Big Rip singularity is considered to be effectively avoided in the quantum analysis. The quantum cosmology of the classical Big Brake singularity was analysed in \cite{Kamenshchik:2007zj}. This is a type of Sudden future singularity where the Hubble rate reaches zero and the deceleration approaches infinity leading to an abrupt brake of the expansion\footnote{In fact, a Sudden singularity occurs at a finite scale factor where the Hubble rate is finite but its cosmic time derivative diverges. Depending on the parameters of the equation of state, this singularity can happen in the future as a Big Brake \cite{Kamenshchik:2007zj} or in the past. This second case, it what we named the \textit{Big D\'emarrage}, it results from a phantom fluid leading to a cosmic expansion that starts with infinite acceleration \cite{Bouhmadi-Lopez:2009pu}.}. It was found that under reasonable assumptions, the DeWitt (DW) criterium is satisfied, i.e. the wave function vanishes in the region of the classical singularity. The \textit{Big D\'emarrage}, another kind of Sudden singularity, and the Big Freeze, the later can be seen as a Big Rip happening at a finite scale factor, were also investigated in the quantum approach based on the WDW equation \cite{Bouhmadi-Lopez:2009pu}. These singularities can result from a dark energy fluid with an appropriate generalised Chaplygin gas equation of state \cite{Mariambigfreeze}. 
In the last mentioned cases (including type IV singularity), the DW criterium for singularity avoidance is satisfied, pointing to a possible avoidance of the singularities \cite{Kamenshchik:2007zj,Bouhmadi-Lopez:2013tua,Bouhmadi-Lopez:2009pu}. An essential result from all these works is that singularity avoidance in quantum cosmology necessarily predict quantum effects at scales much larger than the Planck length.

The heart of quantum cosmology is to apply the quantum theory to the Universe as a whole. This contrasts with the approach in which there is a classical theory with quantum effects (corrections) in certain phases of the evolution. Therefore, quantum cosmology must be based on a quantum theory of gravity. Three major promising approaches are String theory, Loop Quantum Gravity and Quantum Geometrodynamics, where the last two are different developments of Canonical Quantum Gravity \cite{KieferQG,PMonizQC}. Independently of the correct theory of quantum gravity, the quantum cosmology based on the WDW equation should remain valid at least on energies below the Planck scale (if not on all scales) \cite{KieferQG}. The two main challenges of quantum cosmology consist in i) finding a description of the dynamical evolution and ii) in finding the quantum state of the universe with the appropriate initial or boundary conditions. These issues are interrelated although we do not have a robust theory for the boundary conditions or the initial quantum state \cite{KieferQG,PMonizQC,FuturePhscs}. Nevertheless, the WDW equation is crucial for the topic of boundary conditions in quantum cosmology. A relevant feature of the WDW equation is the fact that, for usual scalar fields, it is locally hyperbolic (taking the form of a wave equation) such that it has a well posed initial value problem. This fact is related to the indefinite sign of the kinetic term, which is directly linked to the attractive nature of gravity for usual matter \cite{KieferQG}. The presence of a phantom field changes the structure of this equation, for example, if there is ``phantom dominance'', the WDW equation becomes elliptic (or parabolic), which changes the imposition of the boundary conditions. This is of extreme relevance since while the solutions of hyperbolic equations are ``wave-like", a perturbation of the initial (or boundary) data of an elliptic (or parabolic) equation is felt ``at once'' by essentially all points in the domain. Another important aspect is related to the problem of time: the WDW equation is independent of an external time parameter. On the other hand, the origin of the arrow of time can in principle be related to the structure of this equation \cite{Dabrowski:2006dd,KieferQG}.

In summary, the quantum cosmology of dark energy models leading to classical singularities, is showing very clearly that quantum effects are predicted at scales much larger than the Planck length and that it is possible to find solutions to the WDW equation that could  effectively avoid the classical singularity. Many open questions remain, such as the appropriate boundary conditions, the classical-quantum correspondence, the problem of time and the interpretation of the wave function.

In this work, we apply the methods of quantum cosmology to the LSBR \cite{Bouhmadi-Lopez:2014cca}. 
The quantisation of the classical model presented in \cite{Bouhmadi-Lopez:2014cca} will be carried within 
the geometrodynamical approach given by WDW equation. We consider the case where the LSBR is induced by a perfect fluid, 
in which the scale factor is the single degree of freedom, and also the case where the LSBR is induced by a scalar field with phantom character; therefore in a cosmological model with two degrees of freedom. 
In the canonical quantisation we have $[\hat{a},\hat{\pi_{a}}]\neq0$, where $\hat{\pi}_{a}$ is the 
canonically conjugate momentum of the scale factor \textit{a}, therefore in principle one can choose 
different ``factor-ordering'' in the derivation of the WDW equation \cite{KieferQG}. Accordingly, 
for the case with a perfect fluid we will solve the WDW equation within a WKB approximation, for two 
different factor ordering choices. For the phantom scalar field case, we solve the WDW equation within the
Laplace--Beltrami factor-ordering \cite{KieferQG}, with two degrees of freedom: the scale factor 
and a scalar field, which classically describe the geometry and the matter content respectively. 
The solutions are obtained using a Born--Oppenheimer (BO) approximation. In all the approaches, we impose the DW condition and find that it can be verified which might be an indication that the classical LSBR event can be avoided in the quantum description. As will be pointed in the conclusions, the DW condition on the wave function might not be sufficient to guarantee that all the relevant quantities which diverge in the classical model become finite in the quantum description. One should compute the corresponding expectation values and/or probability amplitudes of the physically relevant quantities to have a complete analysis. In this work, we simply check if the DW condition can be verified. The calculation of the expectation values and the probability amplitudes requires further investigation since it is intimately linked to various open questions in quantum cosmology regarding the Hilbert space structure of the solutions  and the classical-quantum correspondence.   

The structure of the article is as follows: in section II, we review the basic aspects of the classical 
model analysed in \cite{Bouhmadi-Lopez:2014cca} and introduce a phantom scalar field which could induce a 
classical LSBR. Section III includes the quantum analysis for the case of a perfect fluid where the scale 
factor is the only degree of freedom; i.e. the LSBR is induced by a perfect fluid which can be fully 
determined through the scale factor which equally defines the geometry. We review the WDW equation, 
find its solutions using the WKB approximation and verify the DW condition using two different factor 
ordering. In section IV we solve the WDW equation for the phantom scalar field case, using a 
BO approximation and evaluate the asymptotic behaviour of the corresponding solutions to 
see if the DW condition can be satisfied. Finally, in section V we present the conclusions. 
Three appendixes are also included: Appendix A contains some mathematical details of the WKB approximation used in section III and also 
the validity of this method. 
Some detailed calculations related to the Parabolic Cylinder functions
appearing in section IV are relegated to appendix B.
Finally, in appendix C we present the conditions for the validity of the BO approximation used in section IV.

\section{The Little Sibling of the Big Rip event: review}

The classical model leading to the future event we are interested
in was exposed in Ref. \cite{Bouhmadi-Lopez:2014cca}, nevertheless, we present here the basic
features of the classical description relevant for the present work.
We consider the Einstein equations compatible with the cosmological
principle, therefore the Universe is assumed to be homogeneous
and isotropic, i.e. it is well described by a Friedmann--Lema\^{i}tre--Roberston--Walker (FLRW) space-time
metric and the matter content is represented by a perfect fluid.
The Friedmann equation governing the evolution of the scale factor
is
\begin{equation}
H^{2}=\frac{8\pi G}{3}\rho_{\textrm{tot}}-\frac{k}{a^{2}},\label{eq:Friedmm}
\end{equation}
where $k=-1,0,1$ for a hyperbolic, flat or
spherical spatial geometry respectively, $H$ is the Hubble parameter and $\rho_{\textrm{tot}}$ includes different forms
of matter-energy, such as matter (baryonic+dark-matter), radiation
and some form of unknown DE. We disregard the interaction
between these fluids, therefore all obey the conservation equation
$\nabla_{\mu}T^{\mu\nu}=0$, where $T^{\mu\nu}$ is the energy-momentum tensor for each component and $\nabla_{\mu}$ stands for the covariant derivative. Since this implies that for pressureless
matter $\rho_{\textrm{mat}}\sim a^{-3}$, while for radiation we have $\rho_{\textrm{rad}}\sim a^{-4}$,
then the asymptotic behaviour of the scale factor will be governed
by the dark energy component $\rho_{\textrm{de}}$. Note that for the late-time
evolution, we can neglect the curvature term. In fact, for a dominating
form of dark energy which by definition is transgressing the strong energy condition
(SEC: $\rho+3p>0$), the DE density,  $\rho_{\textrm{de}}$ will always dominate over the curvature
term. The cosmological constant transgresses the SEC condition, but there are many other possibilities
coming from a whole range of models. For these type of fluids, the Universe faces an accelerated
expansion, according to the Raychaudhuri equation
\begin{equation}\label{eq:Raychad}
\frac{\ddot{a}}{a}=-\frac{4\pi G}{3}(\rho+3p),
\end{equation}
where $p$ is the pressure and the dot stands for derivative with respect to the cosmic time.
As in Ref. \cite{Bouhmadi-Lopez:2014cca} we assume an equation of state that slightly deviates
from that of a cosmological constant (in the following sections we
drop the ``de'' label)
\begin{equation}\label{eq:eostate}
p=-\rho-\frac{A}{3}.
\end{equation}
We will consider $A$ to be positive and very small such that the
model can mimic closely the $\Lambda$CDM in the past and during the
present \cite{Bouhmadi-Lopez:2014cca}. For this choice, the conservation equation for $\rho(a)$
gives a logarithmic dependence with the scale factor \cite{Bouhmadi-Lopez:2014cca}
\begin{equation}\label{eq:dedensity}
\rho(a)={\Lambda}+A\ln\left(\frac{a}{a_{0}}\right),
\end{equation}
where $a_{0}$ is an integration constant. In \cite{Bouhmadi-Lopez:2014cca}, $a_{0}$ was regarded as the present value of the scale factor such
that $\rho(a_{0})=\Lambda$ mimics a cosmological constant
at present. In this work, although we are using the same notation
as in \cite{Bouhmadi-Lopez:2014cca}, we are interested in the late-time cosmological
evolution, therefore we will consider the solution of the Friedmann
equation for the late-time expansion during a DE dominated
epoch. For the fluid obeying the equation of state (\ref{eq:eostate}),
the scale factor diverges in the infinite future as well as the Hubble
rate and the four-dimensional Ricci curvature, but the derivative
of the Hubble rate does not. This future event is what we refer as
``Little Sibling'' of the Big Rip singularity \cite{Bouhmadi-Lopez:2014cca}.

\begin{figure}[h]
\begin{center}
\includegraphics[width=10cm]{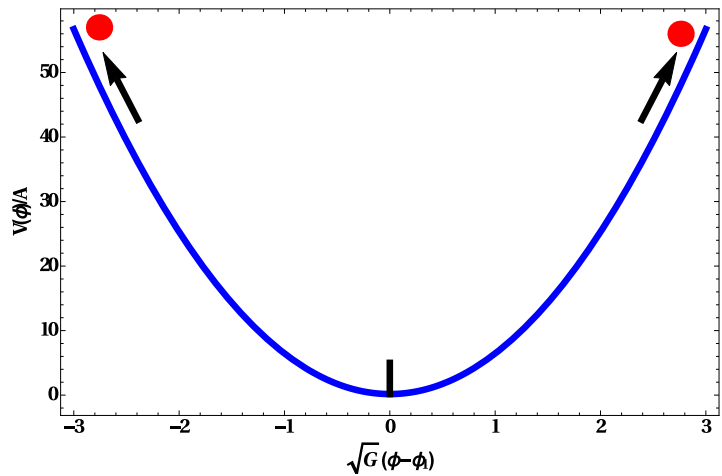}
\end{center}
\caption{Plot of the potential (\ref{eq:potential}). The small vertical line close to the origin corresponds to the initial value $\phi_1$. The abrupt little sibling of the big rip happens at large values as shown by the black arrows. The circles indicated schematically the occurrence of the  little sibling of the big rip.}
\label{plotpotential}
\end{figure}

The dynamics of the previously described perfect fluid can be represented by a scalar field with a phantom character; i.e. 
\begin{equation}
\rho_{\phi}=-\frac{\dot{\phi}^{2}}{2}+V(\phi),\qquad p_{\phi}=-\frac{\dot{\phi}^{2}}{2}-V(\phi),\label{eq:5}
\end{equation}
where $\rho_{\phi}=\rho$ and $p_{\phi}=p$. Using the equation of state (\ref{eq:eostate}) we get
\begin{equation}\frac{d\phi}{da}=\pm{\sqrt{\frac{A}{3}}} \frac{1}{aH(a)},\qquad V(a)={\Lambda}+\frac{A}{6}\left[1+6\ln\left(\frac{a}{a_{0}}\right)\right],\label{potetiala}
\end{equation}
and considering the (asymptotic) Friedmann equation,
\begin{equation}
H^{2}=\frac{8\pi G}{3}\rho,\label{eq:assimptoticfriedmann}
\end{equation}
with $\rho$ defined in Eq. (\ref{eq:dedensity}),
we get two sets of general solutions for the classical trajectory
in configuration space:
\begin{figure}[t]
\begin{center}
\includegraphics[width=10cm]{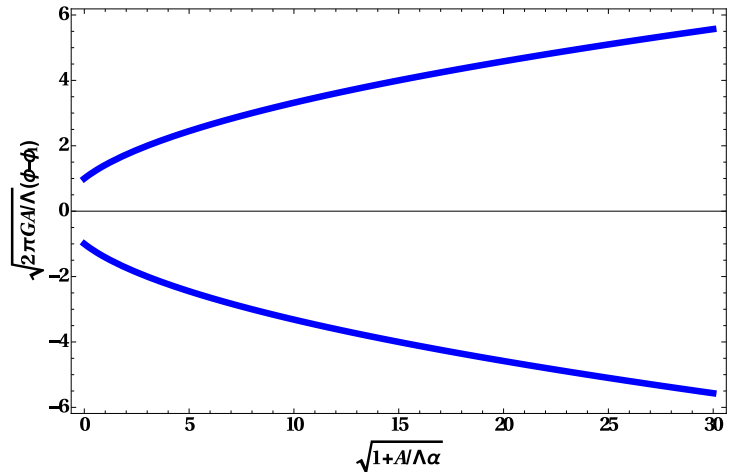}
\end{center}
\caption{Plot of the scalar field versus
the logarithmic scale factor $\alpha=\ln({a}/{a_{0}})$ as given by the expression (\ref{eq:classicaltrajectory}). The little sibling of the big rip is located at large values of $\alpha$. }
\label{plotphia}
\end{figure}
\begin{equation}
\phi(a)=\pm\frac{1}{\sqrt{2\pi G}}\sqrt{\frac{{\Lambda}}{A}+\ln\left(\frac{a}{a_{0}}\right)}+\phi_{1},\label{eq:classicaltrajectory}
\end{equation}
where $\phi_{1}$ is an integration constant fixed as
\begin{equation}
\phi_{1}=\phi(a_{0})\mp\frac{1}{\sqrt{2\pi G}}\sqrt{\frac{\Lambda}{A}}.
\end{equation}
The evolution of the scalar field in terms of $\alpha=\ln\left(\frac{a}{a_{0}}\right)$ is depicted in Fig.~\ref{plotphia}.
Finally, equation (\ref{eq:classicaltrajectory}) can be inverted and through Eq. (\ref{potetiala})  we get
a uni-parametric family of quadratic potentials $V(\phi)$
\begin{equation}\label{eq:potential}
V(\phi)=\frac{A}{6}+2\pi AG\left(\phi-\phi_{1}\right)^{2},
\end{equation}
which is shown in Fig. \ref{plotpotential}.

\section{Quantum study with a perfect fluid}

\subsection{Wheeler--DeWitt equation - review}

The WDW equation can be introduced in the canonical quantisation
of gravity in the so called quantum geometrodynamics approach, we next summarise how it can be obtained under the assumptions of homogeneity and isotropy. From the Einstein--Hilbert gravitational action functional, the Einstein
(Euler--Lagrange) equations can be obtained and used to deduce the Hamiltonian
equations describing the dynamical evolution. It turns out that this
is a constrained dynamical system, in which the constraints
correspond to the invariance of the model with respect to space-time
diffeomorphisms. There are two types of constraints: the momenta constraints
(space diffeomorphisms) and the Hamiltonian constraint (time diffeomorphisms) \cite{KieferQG,PMonizQC}.

For cosmological applications it is usual to consider the principle
of homogeneity and isotropy; i.e, FLRW
space-time metric. The Hamiltonian is derived from the Lagrangian, from which  the dynamical equations can be obtained. In fact, the classical Hamiltonian constraint $\mathcal{H}=0$ corresponds to the Friedmann equation. The passage to the quantum description is done by
promoting all the degrees of freedom and canonically conjugate
momenta to operators acting on some Hilbert space. The usual approach
is to consider the so called minisuperspace, the space of all 3-dimensional
totally symmetric (FLRW) metrics and all matter-energy configurations.
It is analogous to the configuration space in the classical analysis.
The wave function of the universe (in the Schr\"{o}dinger representation) will live on this space and each
point on it represents a FLRW universe with a certain value of the
scale factor and certain well defined values for all the matter-energy
degrees of freedom. The WDW equation governs precisely the distribution of the
wave function in this (quantum) configuration space. It corresponds
to the Hamiltonian constraint (time diffeomorphism invariance) previously
mentioned, where the Hamiltonian operator acting on the wave function
gives zero. Therefore, time is absent in this quantum description and the wave function in the Born interpretation is expected to provide a stationary probability (amplitude) distribution in minisuperspace \cite{{KieferQG}}.  As mentioned in the introduction, in the canonical quantisation procedure
we have to take into account that there are different ``factor-orderings''
in obtaining the WDW equation. 

The action for gravity with a FLRW symmetry can always be written in the following
way \cite{KieferQG,PMonizQC,Albarran:2015tga}
\begin{equation}
S=S_{g}+S_{m},\qquad S_{g}=\frac{3\pi}{4G}\int\left[-\dot{a}^{2}a+ka\right]dt,\label{eq:actionFRW}
\end{equation}
where $S_{m}$ corresponds to the matter part which can be represented
by a perfect fluid dependent on the scale factor (one single degree
of freedom) or include intrinsic matter-energy degrees of
freedom like a scalar field, $\phi$. The corresponding WDW equation can be expressed as
\begin{equation}
\hat{\mathcal{H}}\psi=0,\label{eq:actiongravity}
\end{equation}
where $\hat{\mathcal{H}}$ is the quantum Hamiltonian computed from the above
action.

A reasonable model is one in which the decoherence of the general
superposition of quantum states is absent in those regions of minisuperspace
corresponding to a classically behaving universe \cite{KieferQG}. This requirement
is analogous to the correspondence principle in usual quantum mechanics
where the classical regime is re-obtained for large quantum numbers.
Here, coherence presupposes the possibility for confined wave packet solutions, effectively reproducing the classical trajectory in configuration space. However, in the vicinity of classical singularities,
where the quantum cosmology can have important non-classical effects, decoherence
might be present and the quantum results can differ from the classical description \cite{Dabrowski:2006dd}. Therefore, for any FLRW model with singularities
(or abrupt events such as the LSBR), one should check if
the distribution of the wave function in minisuperspace is such that
these (classical) cosmic events can be effectively avoided with the
quantum approach.

In this section, we study the quantum behaviour of the LSBR induced classically by the perfect fluid with equation of state (\ref{eq:eostate}). We investigate the solutions of the WDW equation and apply the DW criterium for effective singularity avoidance, namely, the condition that the wave function vanishes in the region corresponding to the classical LSBR. We will consider also two different factor ordering choices.

\subsection{Wheeler--DeWitt equation with a perfect fluid - WKB approximation}

We start by considering the simplest case within the WDW equation where the matter content is given by a perfect fluid with a given equation of state, therefore whose energy density is specified by the scale factor. The scale factor
$a$ is then the only canonical variable and the Lagrangian for FLRW
models can be written in the following way \cite{Albarran:2015tga,KieferQG,PMonizQC,BouhmadiLopez:2004mp,BouhmadiLopez:2006pf}
\begin{equation}
L=-\frac{3\pi}{4G}\left[\dot{a}^{2}a-ka\right]-2\pi^{2}a^{3}\rho,
\end{equation}
where $\rho$ is the total energy density and the dot represents derivative
with respect to the cosmic time.
Here we are neglecting the cosmological constant (which can be incorporated
in the dark-energy density). The canonical conjugate momentum of the
scale factor is
\begin{equation}\label{canonical momentum}
\pi_{a}\equiv\frac{\partial L}{\partial\dot{a}}=-\frac{3\pi }{2G}a\dot{a},
\end{equation}
and introducing the classical Hamiltonian, $\mathcal{H}\equiv\dot{a}\pi_{a}-L$, the corresponding
constraint $\mathcal{H}=0$ is given by
\begin{equation}\label{eq:wkbHamiltconstraint}
-\frac{G}{3\pi}\frac{\pi_{a}^{2}}{a}-\frac{3\pi}{4G}ka+2\pi^{2}a^{3}\rho=0.
\end{equation}
This equation can be simplified, yielding precisely the Friedmann equation (\ref{eq:Friedmm}). As for the quantum model, depending on the factor ordering, we will get different equations. We will verify if the DW condition \cite{KieferQG,PMonizQC}, i.e the vanishing of the wave function when the scale factor goes to infinity for a  $\rho$ given by Eq. (\ref{eq:dedensity}), can be satisfied and if it depends on the factor ordering choice.

\subsubsection{First quantisation procedure ($a\mathcal{\hat{H}}(a,\hat{\pi}_{a})\psi(a)=0$)}

The simplest way to obtain the WDW equation with the scale factor as the single degree of freedom is to multiply the Hamiltonian constraint, Eq. (\ref{eq:wkbHamiltconstraint}), by $a$ and make the substitution $\pi_{a}\rightarrow\hat{\pi}_{a}=-i\hbar\frac{d}{da}$. The corresponding WDW equation is
\begin{equation}\label{firstquant}
\left[\hbar^{2}\frac{G}{3\pi}\frac{d^{2}}{da^{2}}+V(a)\right]\psi(a)=0,
\end{equation}
where
\begin{equation}
V(a)=-\frac{3\pi}{4G}ka^{2}+2\pi^{2}a^{4}\rho(a).
\end{equation}
For the case of a DE fluid obeying the equation of state
(\ref{eq:eostate}), the wave function of the universe obeys the following
differential equation
\begin{equation}
\left\{\hbar^{2}\frac{G}{3\pi}\frac{d^{2}}{da^{2}}-\frac{3\pi}{4G}ka^{2}+2\pi^{2}a^{4}\left[{\Lambda}+A\ln\left(\frac{a}{a_{0}}\right)\right]\right\}\psi(a)=0,
\end{equation}
and for $k=0$, it reduces to
\begin{equation}\label{eq:wdwWKB1}
\left\{\frac{d^{2}}{da^{2}}+\frac{6\pi^{3}}{\hbar^{2}G}a^{4}\left[{\Lambda}+A\ln\left(\frac{a}{a_{0}}\right)\right]\right\}\psi(a)=0.
\end{equation}
We perform the following changes to achieve a dimensionless WDW equation,
\begin{equation}\label{eq:dimss1}
 u\equiv\frac{a}{a_0},\ \ \Omega_{{\Lambda}}\equiv\frac{8 \pi G}{3H_0^2}{\Lambda}, \ \ \Omega_{A}\equiv\frac{8 \pi G}{3H_0^2}A, \ \ \eta\equiv\frac{\pi  a_0^3H_0}{G\hbar}.
\end{equation}
Therefore, the equation (\ref{eq:wdwWKB1}) becomes
\begin{equation}\label{eq:wdwWKBia}
\left\{\frac{d^{2}}{du^{2}}+\left(\frac{3}{2}\eta\right)^2u^{4}\left[\Omega_{{\Lambda}}+\Omega_{A}\ln(u)\right]\right\}\psi(u)=0.
\end{equation}
Using the WKB (Wentzel, Kramers, Brillouin) approximation method at first order, which is enough to ensure that the DW condition is satisfied; i.e. $\Psi(u)$ vanishes for $u\rightarrow\infty$, the wave function of the universe can be described by \cite{Mathew,Albarran:2015tga}
\begin{equation}\label{eq:solaH}
\psi(u)\approx\sqrt{\frac{2}{3\eta}\frac{1}{u^2}}\left[\Omega_{{{\Lambda}}}+\Omega_{A}\ln(u)\right]^{-\frac{1}{4}}\left\{C_1e^{iS_0(u)}+C_2e^{-iS_0(u)}\right\},
\end{equation}
where $C_1$ and $C_2$ are constants and the function $S_0(u)$ can be written according to Eq. (\ref{eq:erfiintegralim}) as
\begin{equation}\label{S0u}
S_0(u)\approx\eta\frac{u^{3}}{2}\left[\Omega_{{{\Lambda}}}+\Omega_{A}\ln(u)\right]^{\frac{1}{2}}.
\end{equation}

\subsubsection{Second quantisation procedure ($\mathcal{\hat{H}}(a,\hat{\pi}_{a})\psi(a)=0$)
- Laplace--Beltrami factor ordering}

The Laplace--Beltrami factor ordering is perhaps better motivated for two or higher dimensional WDW equations; i.e. with two or more physical degrees of freedom, since it is based on a covariant generalisation of the Laplacian operator in minisuperspace \cite{KieferQG}. 
The starting point for the Laplace--Beltrami factor ordering for the case of a perfect fluid specified by the scale factor is to consider the wave equation obtained
from the classical Hamiltonian constraint, Eq. (\ref{eq:wkbHamiltconstraint}),
without multiplying by $a$ (c.f. also, for example,  Ref. \cite{Albarran:2015tga}). Therefore,
\begin{equation}\label{eq:waveeq}
\left[-\frac{G}{3\pi}\frac{\hat{\pi}_{a}^{2}}{a}+2\pi^{2}a^{3}\rho\right]\psi(a)=0,
\end{equation}
here we are already restricting the analysis to a flat geometry
($k=0$). Introducing the change of variable
\begin{equation}\label{eq:newvar2}
x\equiv\left(\frac{a}{a_{0}}\right)^{\frac{3}{2}},
\end{equation}
such that
\begin{equation}\label{eq:classhamil}
\frac{\hat{\pi}_{a}^{2}}{a}=-\hbar^{2}\left[a^{-\frac{1}{2}}\frac{d}{da}\right]\left[a^{-\frac{1}{2}}\frac{d}{da}\right]=-\frac{9}{4}\frac{\hbar^{2}}{a_{0}^{3}}\frac{d^{2}}{dx^{2}},
\end{equation}
the expression (\ref{eq:waveeq}) can be rewritten as
\begin{equation}
\left[\frac{d^{2}}{dx^{2}}+\tilde{V}(x)\right]\psi(a)=0.
\end{equation}
Using the definitions of the parameters realised in Eq. (\ref{eq:dimss1}), the effective potential $\tilde{V}(x)$ is the following 
\begin{equation}\label{eq:PotWKB2}
\tilde{V}(x)\equiv {\eta}^{2}x^{2}\left[\Omega_{{\Lambda}}+\frac{2}{3}\Omega_{A}\ln(x)\right].
\end{equation}
The general solution in the WKB, first order approximation, in this case also verifies the DW condition 
\begin{equation}\label{eq:solLB}
\psi(x)\approx\frac{1}{\sqrt{\eta}}\left\{x^{2}\left[\Omega_{{\Lambda}}+\frac{2}{3}\Omega_{A}\ln(x)\right]\right\}^{-\frac{1}{4}}\left\{\tilde{C_1}e^{i\tilde{S_0}(x)}+\tilde{C_2}e^{-i\tilde{S_0}(x)}\right\},
\end{equation}
where $\tilde{C_1}$ and $\tilde{C_2}$ are constants and the function $\tilde{S_0}(x)$ can be written according to Eq. (\ref{eq:erfiintegralim}) as
\begin{equation}
\tilde{S_0}(x)\approx\eta\frac{x^{2}}{2}\left[\Omega_{{{\Lambda}}}+\Omega_{A}\ln(x)\right]^{\frac{1}{2}}.
\end{equation}

\section{Quantum study with a phantom scalar field}

In this section, we study the quantum behaviour of the LSBR 
using an approach based on considering a phantom scalar field. 
We consider the Born--Oppenheimer (BO) approximation for the the WDW equation and, as in the
previous section, we check the fulfilment of the DW criterium as an indication of singularity avoidance.

\subsection{Wheeler--DeWitt equation with a phantom field - Review}

Assuming FLRW symmetry the gravitational action with a minimally coupled phantom scalar field 
(for vanishing cosmological constant) can be written as (c.f. Ref. \cite{Dabrowski:2006dd}):
\begin{equation}\label{eq:actionFRWphantom}
S=\frac{3\pi}{4G}\int\left[-\dot{a}^{2}a+ka\right]dt-\pi^{2}\int a^{3}\left[\dot{\phi}^{2}+2V(\phi)\right]dt.
\end{equation}
The corresponding Wheeler-DeWitt
equation $\hat{\mathcal{H}}\psi(\alpha,\phi)=0$, for flat spatial geometry is \cite{Dabrowski:2006dd}
\begin{equation}\label{eq:wdweqBornOp}
\frac{\hbar^{2}}{2}\left[\frac{\kappa^{2}}{6}\frac{\partial^{2}}{\partial\alpha^{2}}+\frac{\partial^{2}}{\partial\phi^{2}}\right]\psi(\alpha,\phi)+a_{0}^{6}e^{6\alpha}V(\phi)\psi(\alpha,\phi)=0,
\end{equation}
where $\alpha\equiv\ln\left({a}/{a_{0}}\right)$, $\kappa^{2}\equiv8\pi G$, $V(\phi)$ 
is the potential for the scalar field, which in our model is quadratic, Eq. (\ref{eq:potential}),
and we have taken $k=0$. 
Here we have used the Laplace--Beltrami factor-ordering since, as discussed in section III,
in two dimensions it leads to the Laplace--Beltrami operator which is the covariant generalisation of 
the Laplacian operator in minisuperspace \cite{KieferQG}. 
Apart from this formal motivation, we restrict to the consideration of this 
factor ordering because we have seen on the previous section that a change on the factor ordering seems to leave unaffected the conclusion about LSBR avoidance.

Before proceeding to solve the WDW equation some comments are in order.
In the first place, one should keep in mind that throughout this paper the validity of the quantum 
description based on the WDW equation is limited to a subset of the minisuperspace corresponding to the 
validity of the DE dominance assumption ($\rho_{\rm tot}\approx\rho_{\rm de}=\rho$ in Eq.~(\ref{eq:Friedmm})). 
This implies that we are considering through the paper the existence of a finite minimum for $\alpha$, 
$\alpha_{\rm min}>-{{\Lambda}}/A$, 
as can be obtained by considering the positivity of the energy density given by Eq.~(\ref{eq:dedensity}).
In the second place, $\alpha$ and $\phi$ are not independent classically, according to 
Eq.~(\ref{eq:classicaltrajectory}). 
Naively, this could seem to imply that to check the fulfilment of the DW criterium we
should search for solutions $\psi(\alpha,\phi)$ that decay along that line $\phi=\phi^{\rm class}(\alpha)$,
$\psi(\alpha,\phi^{\rm class}(\alpha))\rightarrow0$,
when the scale factor goes to infinity. Nevertheless, when the energy density
approaches a certain quantum gravity threshold, the quantum effects
should become important breaking the classical
constraint $\phi=\phi^{\rm class}(\alpha)$. Indeed, although the potential is classically
obtained in section II, $\alpha$ and $\phi$ are independent in the quantum description. 
Therefore, we must investigate the evolution of the wave function $\psi(\alpha,\phi)$
for $\alpha\rightarrow\infty$ and arbitrary values of $\phi$.
Finally, for a compatible link between the quantum solutions and the classical behaviour, 
far from the quantum effects dominance, 
the wave function should be described by wave-packets centered in each point of the classical trajectory 
$\phi=\phi^{class}(\alpha)$ \cite{KieferQG}. 
We will not consider this issue in the present work.

\subsection{Born-Oppenheimer approximation}

To solve the WDW equation (\ref{eq:wdweqBornOp}) we will now adopt the BO
approximation which considers that the wave function can be factorised into two parts,
corresponding to the geometric and matter parts. This is
\begin{equation}
\psi_{k}(\alpha,\phi)=\varphi_{k}(\alpha,\phi)C_{k}(\alpha).\label{eq:BOapr}
\end{equation}
The matter part $\varphi(\alpha,\phi)$ is assumed to satisfy the
hypothesis of adiabatic dependence with the scale factor, such that
the terms containing its derivatives with respect to $\alpha$ can be neglected. 
The consideration of the BO ansatz (\ref{eq:BOapr}) in the WDW equation (\ref{eq:wdweqBornOp}) 
leads to two differential equations, 
one for the geometric part of the wave function and the other for the matter part.
Those are
\begin{equation}\label{eq:WSWmatter}
\frac{\hbar^{2}}{2}\frac{\partial^{2}\varphi_{k}(\alpha,\phi)}{\partial\phi^{2}}+\left[a_{0}^{6}e^{6\alpha}V(\phi)-E_{k}(\alpha)\right]\varphi_{k}(\alpha,\phi)=0,
\end{equation}
\begin{equation}\label{eq:wdwgeometric}
\frac{\kappa^{2}}{6}\frac{\partial^{2}C_{k}(\alpha)}{\partial\alpha^{2}}+k^{2}(\alpha)C_{k}(\alpha)=0,
\end{equation}
where $k^{2}(\alpha)\equiv\frac{2}{\hbar^{2}}E_{k}(\alpha)$.  In this approximation, the matter part of the wave function has no backreaction in the geometric part. The two equations become decoupled and the matter part has only an indirect influence on the geometric part through the eigenvalues $E_{k}(\alpha)$. The
validity of this Born-Oppenheimer approximation is discussed in appendix C.

\subsubsection{Solving the matter part}

Substituting the potential given by Eq.~(\ref{eq:potential}), 
in the matter part of the WDW equation, Eq.~(\ref{eq:WSWmatter}),
we obtain
\begin{equation}\label{eq:WSWmatterpotential}
\frac{\hbar^{2}}{2}\frac{\partial^{2}\varphi_{k}(\alpha,\phi)}{\partial\phi^{2}}+
\left\{a_{0}^{6}e^{6\alpha}\left[\frac{A}{6}+2\pi G A\left(\phi-\phi_{1}\right)^{2}\right]-
E_{k}(\alpha)\right\}\varphi_{k}(\alpha,\phi)=0.
\end{equation}
This equation is not analogous to the quantum harmonic oscillator but is similar to a repulsor instead. 
This is a consequence of considering a phantom scalar field and rather than a usual (quintessence) scalar 
field.
Therefore, the spectrum is continuous in this case and we can consider that
$E_{k}(\alpha)=E_{k}$ and $k^{2}(\alpha)\equiv k^{2}$ are continuous parameters independent of $\alpha$,
where $k$ is not restricted to real values.  
The solutions of Eq.~(\ref{eq:WSWmatterpotential}) are (see appendix B for details)
\begin{equation}\label{sol1}
 \varphi_{k}^{(1,2)}(\alpha,\phi)=W\left(\beta,\,\pm z\right),
\end{equation}
\begin{equation}\label{sol2}
 \varphi_{k}^{(3)}(\alpha,\phi)=K^{-1/2} W\left(\beta,\, z\right)+iK^{1/2} W\left(\beta,\, -z\right),
\end{equation}
and 
\begin{equation}\label{sol3}
 \varphi_{k}^{(4)}(\alpha,\phi)=K^{-1/2} W\left(\beta,\, z\right)-iK^{1/2} W\left(\beta,\, -z\right),
\end{equation}
where $W\left(\beta,\, z\right)$ is a parabolic cylinder function \cite{Abramow},
$K=\sqrt{1+e^{2\pi\beta}}-e^{\pi\beta}$,
\begin{equation}
 \beta=-\frac{1}{2\hbar(\pi G)^{1/2}}\left[\frac{A^{1/2}a_0^3e^{3\alpha}}{6}-
 \frac{E_k}{A^{1/2}a_0^3e^{3\alpha}}\right],
\end{equation}
and
\begin{equation}
 z=\frac{2 a_0^{3/2} e^{3\alpha/2} (\pi G A)^{1/4}}{\hbar^{1/2}}(\phi-\phi_1).
\end{equation}
Thus, the matter part of the wave function is a combinations of these solutions \cite{Abramow}
\begin{equation}
  \varphi_{k}(\alpha,\phi)=c_1  \varphi_{k}^{(1)}(\alpha,\phi)+c_2 \varphi_{k}^{(2)}(\alpha,\phi)+c_3 \varphi_{k}^{(3)}(\alpha,\phi)+c_4 \varphi_{k}^{(4)}(\alpha,\phi),
\end{equation}
with $c_i$'s constants. As we are dealing with a second order differential equation only at most two of the previous functions are linearly independent.

As shown in appendix B, in the limit $\alpha\rightarrow\infty$ these
parabolic cylinder functions go as
\begin{equation}
 W\left(\beta,\, z\right)\sim e^{-3\alpha/4}\cos(e^{3\alpha}),
\end{equation}
and
\begin{equation}
 W\left(\beta,\, -z\right)\sim e^{-3\alpha/4}\sin(e^{3\alpha}),
\end{equation}
Therefore, the matter part of the wave function tend to zero when $\alpha\rightarrow\infty$ for arbitrary $c_i$'s. 
This imply that the DW condition is satisfied
if the geometric part of the wave function stays finite in this limit.

\subsubsection{Solutions to the geometric part}

The solutions of equation (\ref{eq:wdwgeometric}) are oscillatory
for $E_{k}>0$,
\begin{equation}
C_{k}(\alpha)=a_{1}e^{i\frac{\sqrt{12E_{k}}}{\kappa\hbar}\alpha}+a_{2}e^{-i\frac{\sqrt{12E_{k}}}{\kappa\hbar}\alpha},
\end{equation}
and have an exponential behaviour for $E_{k}<0$
\begin{equation}
C_{k}(\alpha)=b_{1}e^{\frac{\sqrt{12\left\vert E_{k}\right\vert}}{\kappa\hbar}\alpha}+b_{2}e^{-\frac{\sqrt{12\left\vert E_{k}\right\vert}}{\kappa\hbar}\alpha},
\end{equation}
where $a_{1}, a_{2}, b_{1},b_{2}$ are constants. Therefore, to make sure that DW criterion for singularity 
avoidance is  fulfilled we impose $b_1=0$.

\section{Conclusions and Outlook}

The main purpose of the present work was to analyse the behaviour of the Universe when approaching a LSBR \cite{Bouhmadi-Lopez:2014cca} within the framework of quantum cosmology \cite{KieferQG,PMonizQC}.
We have obtained the WDW equation in the context of quantum geometrodynamics
for two cases: i) a perfect fluid with the equation of state (\ref{eq:eostate}) and ii) a dominant scalar field of phantom character whose potential is given in Eq. (\ref{eq:potential}). We applied the DW criterium for singularity avoidance, namely the condition that the wave function should vanish in the region of the classical singularity. 
In the first case, the DW condition is satisfied, independently of the two factor ordering
considered. This can easily be verified by the solutions obtained in the WKB approximation, Eqs. (\ref{eq:solaH}) and (\ref{eq:solLB}). For two degrees of freedom, we have also shown that it is possible to find solutions obeying the DW condition for any fixed value of the phantom field $\phi$ or even
along the classical trajectory, c.f. 
Eqs. (\ref{sol1}), (\ref{sol2}), and (\ref{sol3}). These results might point to a possible resolution of the LSBR in the quantum description.

For the scalar field case, we used a BO approximation leading to two separate equations, one for 
the geometric part of the wave function and the other for the matter part. In this approximation, 
the matter part of the wave function has no backreaction in the geometric part. The two equations 
become decoupled and the matter part has only an indirect influence on the geometric part through 
the eigen values $E_{k}$. The solutions for the geometric part can be divergent (which we disregard) 
and convergent exponentials or oscillatory for $E_{k}<0$ or $E_{k}>0$, respectively.
Due to the phantom character of the scalar field, the equation for the matter part is analogous to a quantum harmonic 
repulsor. 
The solutions of this equation are Parabolic Cylinder functions \cite{Abramow}.
As it is shown in appendix B in detail, these solutions vanish in the limit of arbitrarily large scale 
factor. Therefore, the DW condition is asymptotically satisfied.

The fundamental starting point for the classical model with the LSBR is the equation of state (\ref{eq:eostate}), slightly different from that of a cosmological constant. For the case of a perfect fluid, this relation, which is implicit in the energy density, is directly inserted in the WDW equation in the matter component. Since in this case the matter component has no implicit degrees of freedom, one can say that the traces of the LSBR equation of state are fully encapsulated in the geometric part of the WDW. On the other hand, for the phantom scalar field case, the WDW is two-dimensional and the LSBR signature can be inserted on the phantom part via the scalar field potential. 

As mentioned in the introduction, the issue of boundary or initial conditions for the wave function is of extreme relevance in quantum cosmology and it still is in some sense an open question \cite{KieferQG,PMonizQC}. The no-boundary wave function, the Tunnelling wave functions, were essentially motivated by the application of quantum cosmology to the very early universe (see \cite{FuturePhscs} and references therein). The no-boundary proposal was in some sense unique since it tried to bring together the dynamics and the problem of the initial state. Nevertheless, so far there is no unified theory of boundary conditions and dynamics from which that relation can be derived \cite{FuturePhscs}. A more general suggestion, the one we used in this work, was first given by DeWitt, in what has been named the DeWitt condition. It is a general condition imposed on the wave function saying that it should be zero in the regions corresponding to the classical singularities. Again, at the present moment there is no fundamental proof based on quantum gravity that the solutions to the WDW equation should obey this condition \cite{FuturePhscs}. If this is found then one could say that the theory excludes the (geometric) singularity problems without any ambiguity. 

For genuinely quantum regions, the
correspondence  between the classical and quantum predictions should
in principle break down and these are expected to occur at the vicinity
of classical singularities. Now, as pointed in the introduction, one way
for this to happen is to get wave packets solutions (following the
classical trajectory) that become smeared out in the classical singularity
region, effectively avoiding it \cite{KieferQG,Dabrowski:2006dd} and another way is to have a decaying wave function in those regions, satisfying the DW condition \cite{KieferQG}. The DW condition, taken as a criterium to avoid the classical singularities is based on the existence of square integral functions, and therefore on a consistent probability interpretation for the wave function. Accordingly the probability
amplitudes for wave packets should vanish in the limit when the variables
(of the representation) go to infinity. The problem is the fact
that these square integral functions require an appropriate Hilbert
space and it is not obvious
that this can always be done in the quantum cosmology based on the
WDW equation (c.f. for example the references \cite{Barvinsky:2013aya,Kamenshchik:2013naa,Kamenshchik:2012ij,Barvinsky:1993jf} for cases where this is doable).

The LSBR is an event happening in the limit when
the scale factor goes to infinity, leading to the divergence of the
Ricci parameter, the Hubble parameter and of course the dominant dark
energy density. Can we avoid, with the quantum treatment, the region
of configuration space corresponding to the divergent scale
factor? As mentioned, in this work we tried to impose the DW condition on the solutions to the
WDW equation in order to achieve this goal. What about the other quantities
mentioned: how to compute them in the quantum geometrodynamics
approach? In the quantum description, cosmological quantities such
as the scale factor, the Hubble parameter (and its derivatives), or
the Ricci curvature are represented by operators. Therefore, in principle if we
have a normalised wave function in minisuperspace we can evaluate the expectation
values and probability densities for these observables.
Again, we emphasise the fact that the interpretation of the results for the expectation values in the
quantum cosmology based on the WDW equation depends crucially on a
consistent probability interpretation for the wave function and therefore on a minisuperspace
with a proper Hilbert space nature, which remains an open question. Strictly speaking, the DW condition on the wave function is not sufficient to guarantee that all the relevant quantities which diverge in the classical model become finite in the quantum description. One should compute the corresponding expectation values and/or probability amplitudes to have a complete analysis. In this work, we simply checked if the DW condition could be verified. The calculation of expectation values and probability amplitudes requires further investigation since it is intimately linked to various open questions in quantum cosmology regarding a Hilbert space structure and the classical-quantum correspondence.   

Nevertheless, it can be shown for the case with the scale factor as the single degree of freedom, for solutions Eqs.(\ref{eq:solaH}) and (\ref{eq:solLB}) obtained in the WKB method, that the energy density expectation value is finite. If the effective Friedmann equation, obtained by replacing the classical quantities by the corresponding expectation values, remains valid, it would then imply that the Hubble rate expectation value is also finite in the quantum description. Nevertheless, this is not completely clear and requires further investigation, first of all because the calculation of the expectation value of the Hubble parameter depends on the factor ordering and second, because this effective Friedmann equation might be only valid for the classical regime.  In fact, the standard quantum cosmology implied here,
based on the WDW equation, is strictly speaking a canonical quantisation of the classical theory of GR. In the genuine quantum regions mentioned
the classical gravitational theory of GR is expected to break and we might
need a different theory of quantum gravity, although as it was mentioned in the introduction, many results regarding the WDW analysis of late time cosmology are expected to hold \cite{Dabrowski:2006dd}. Does the (effective) Friedman equation $
<\psi\left|\hat{H}^{2}(\hat{a},\hat{\pi}_{a})\right|\psi>=\frac{8\pi G}{3}<\psi\left|\hat{\rho}(\hat{a})\right|\psi>$,
using the solution to the WDW equation, correspond exactly to the
classical one? If not, is there any region in minisuperspace in which
it is valid, in accordance with the correspondence principle?  
Notice that the above expression contrasts with the semi-classical approach in
which no quantisation is done in the geometric side of the Einstein
equations, $R_{\mu\nu}-\frac{1}{2}g_{\mu\nu}R=\kappa^{2}<\varPsi\left|\hat{T}_{\mu\nu}\right|\varPsi>$ (c.f. for example Ref. \cite{Kuo:1993if}). Historically this approach has been comparatively more explored, in which the energy momentum is replaced by the expectation value computed from the quantum state describing the matter, where $\varPsi$ is the wave function representing the degrees of freedom of some quantum field theory in the fixed background of curved space-time. These issues regarding the calculation of cosmological quantities in the quantum description and their relation to the classical model require further investigation.

Regarding the topic of future cosmic singularities,
there are many open questions. First of all, what kind of observational predictions could distinguish
between the different models with late-time acceleration and dark energy related singularities? Another interesting aspect implicit at least in some of these models is the possibility of matter-energy transgressing most or even all of the classical energy conditions. This lead to the idea that these conditions can be refined through semi-classical or purely quantum arguments \cite{Martin-Moruno:2013wfa,Martin-Moruno:2013sfa,Bousso:2015wca}. A new door on the physics of phase transitions might be open if these non-linear quantum energy conditions (and their interpretations) are profoundly explored.

\section*{Acknowledgments}

IA is supported by the Portuguese Agency ``Funda\c{c}\~{a}o para a Ci\^{e}ncia e Tecnologia" through the Grant PTDC/FIS/111
032/2009. The work of MBL is supported by the Portuguese Agency ``Funda\c{c}\~{a}o para a Ci\^{e}ncia e Tecnologia" through an Investigador FCT Research contract, with reference IF/01442/2013/CP1196/CT0001. She also wishes to acknowledge the support from the Portuguese Grants PTDC/FIS/111032/2009 and UID/MAT/00212/2013  and the partial support from the Basque government Grant No. IT592-13 (Spain) and the  Spanish Grant No. FIS2014-57956-P. FC is supported by the Portuguese Agency ``Funda\c{c}\~{a}o para a Ci\^{e}ncia e Tecnologia" through the Grant PTDC/FIS/111032/2009. PMM acknowledges financial support from the Spanish Ministry of Economy and Competitiviness through the postdoctoral training contract FPDI-2013-16161, and the project FIS2014-52837-P.

\appendix

\section{The WKB approximation}

\subsection{general description}

For the following second order differential equation 
\begin{equation}\label{eq:wdwrWKBia}
\left[\frac{d^{2}}{dy^{2}}+V(y)\right]\psi(y)=0,
\end{equation}
where
\begin{equation}\label{eq:Vr}
V(y)\equiv \tilde{\eta}^2 y^{2s}\left[\Omega_{\Lambda}+b\ln(y)\right],
\end{equation}
the lowest order in the  WKB approximation gives \cite{Mathew}
\begin{equation}
\psi(y)\approx B_{1}e^{iS_0(y)}+B_{2}e^{-iS_0(y)},
\end{equation}
where $B_1$ and $B_2$ are constants and
\begin{equation}\label{Soy}
S_0(y)=\int_{y_1}^{y}{\sqrt{V(y)}dy}.
\end{equation}
The variable $y$ can be written as a positive power of $a$, according to (\ref{eq:dimss1}) and (\ref{eq:newvar2}). Therefore, when  $a\rightarrow\infty$, $y\rightarrow\infty$. Since $V(y)$ is always positive, the solutions for the zeroth order WKB approximation are purely oscillatory. The lower order WKB approximation do not fulfill the DW condition, so we go to next order. 

The WKB approximation in the first order will allow us to have an approximate evaluation of the asymptotic behavior of the independent solutions to our wave equation. the general expression is
\begin{equation}\label{eq:wkbfirstorder}
\psi(y)\approx\left[-V(y)\right]^{-\frac{1}{4}}\left[B_{1}e^{iS_0(y)}+B_{2}e^{-iS_0(y)}\right].
\end{equation}
As we can see, for large values of $y$,  the function $V(y)$ diverges and therefore, the wave-function, $\psi(y)$ vanishes.  The DW condition is satisfied and the Little Sibling event is effectively avoided in the quantum description with one degree of freedom, independently of the two factor orderings considered here. Only the parameters $s$, $b$ and $\tilde{\eta}$ depend on the factor ordering:
\begin{enumerate}
\item For the first quantization procedure used in the present work, we use the operator $a\mathcal{\hat{H}}(a,\hat{\pi}_{a})$. Then, the parameters above are written as follows
\begin{equation}\label{defpams2}
s=2, \ \ b=\Omega_A, \ \ \tilde{\eta}=\frac{3\pi a_0^3H_0}{2G \hbar}.
\end{equation}
\item For the second factor ordering choice, which corresponds to the operator $\mathcal{\hat{H}}(a,\hat{\pi}_{a})$, the parameters are written as
\begin{equation}\label{defpams1}
s=1, \ \ b=\frac{2}{3}\Omega_A, \ \ \tilde{\eta}=\frac{\pi a_0^3H_0}{G \hbar}.
\end{equation}
\end{enumerate}
The integral in Eq. (\ref{Soy}) is given by
\begin{equation}\label{s01}
S_0(y)\equiv\int_{y_1}^{y} \tilde{\eta}y^{s}\left[\Omega_{\Lambda}+b\ln(y)\right]^{\frac{1}{2}}dy,
\end{equation}
after performing an integral by parts we get
\begin{eqnarray}\label{eq:erfiintegral}
S_0(y)=\left.\tilde{\eta}\frac{y^{s+1}}{s+1}\left[\Omega_{\Lambda}+b\ln(y)\right]^{\frac{1}{2}}\right\vert_{y_1}^y-\int_{y_1}^{y}\tilde{\eta}\frac{by^{s}}{2\left(s+1\right)}\left[\Omega_{\Lambda}+b\ln(y)\right]^{-\frac{1}{2}}dy.
\end{eqnarray}

The integral of the second term can be done defining an auxiliary variable 
$t=i\sqrt{\frac{s+1}{b}\left[\Omega_\Lambda+b\ln(y)\right]}$. This leads to
\begin{eqnarray}\label{eq:erfiintegral2}
S_0(y)=\tilde{\eta}\frac{y^{s+1}}{s+1}\left[\Omega_{\Lambda}+b\ln(y)\right]^{\frac{1}{2}}-\tilde{\eta}\frac{\sqrt{\pi b}e^{-\frac{\Omega_{\Lambda}}{b}\left(s+1\right)}}{2\left(s+1\right)^{\frac{3}{2}}}\left\{\textrm{erfi}\left[\left(\frac{s+1}{b}\right)^{\frac{1}{2}}\sqrt{\Omega_{\Lambda}+b\ln(y)}\right]\right\}\\ \nonumber
-\tilde{\eta}\frac{y_1^{s+1}}{s+1}\left[\Omega_{\Lambda}+b\ln(y_1)\right]^{\frac{1}{2}}+\tilde{\eta}\frac{\sqrt{\pi b}e^{-\frac{\Omega_{\Lambda}}{b}\left(s+1\right)}}{2\left(s+1\right)^{\frac{3}{2}}}\left\{\textrm{erfi}\left[\left(\frac{s+1}{b}\right)^{\frac{1}{2}}\sqrt{\Omega_{\Lambda}+b\ln(y_1)}\right]\right\},
\end{eqnarray}
where $y_1$ is an arbitrary integration constant such that $a_1$ (corresponding to $y_1$) is larger than $a_0$. 
The function $\textrm{erfi}(z)$ is the ``imaginary error function'', which is related with the error function (defined
by Eq.~7.1.1 of page 297 in Ref.~\cite{Abramow}) through ${\rm erfi}(z)=-i{\rm erf}(iz)$.
The second term on the rhs in Eq.~(\ref{eq:erfiintegral2}) is much smaller than the first term for large values of 
$y$ (as it can be noted considering Eq.~7.1.23 of page 298 in Ref.~\cite{Abramow}), and
the third and four terms can, of course, be dismissed in this limit. Therefore, we can neglect it for big values of $y$.
\begin{equation}\label{eq:erfiintegralim}
S_0(y)\approx\tilde{\eta}\frac{y^{s+1}}{s+1}\left[\Omega_{\Lambda}+b\ln(y)\right]^{\frac{1}{2}}.
\end{equation}

This approximation (\ref{eq:erfiintegralim}) is then applied in equations (\ref{S0u}) and (\ref{eq:solLB}), where the value of the parameters are written in (\ref{defpams2}) and (\ref{defpams1}), for the first and the second quantisation procedures, respectively.

\subsection{Validity of the WKB approximation.}

The wave equations we have obtained in Eqs. (\ref{eq:solLB}) and (\ref{eq:solaH}) using the two different factor-orderings
are both of the form
\begin{equation}
\left[\partial_{y}^{2}+\tilde{\eta}^{2}g(y)\right]\psi(y)=0,\label{eq:wkbformal}
\end{equation}
for equations of the form (\ref{eq:wkbformal}), the validity of the WKB method in the zeroth order approximation; i.e. the solutions (\ref{eq:solaH}) and (\ref{eq:solLB}) without the global multiplicative prefactors, is ensured if
\begin{equation}
\frac{1}{\tilde{\eta}}\left|\frac{1}{2g(y)^{\frac{3}{2}}}\frac{dg(y)}{dy}\right|\ll1.
\end{equation}
On the other hand, for the first order approximation the inequality that must be satisfied is 
\begin{equation}
\frac{1}{\tilde{\eta}^{2}}\left|\frac{5\dot{g}^{2}(y)-4\ddot{g}(y)g(y)}{16g^{3}(y)}\right|\ll1,
\end{equation}
where dot stands  for a derivative respect to $y$. This condition must be fulfilled for the validity of the equations (\ref{eq:solaH}) and (\ref{eq:solLB}). These inequalities are  obtained as conditions in order to neglect the remaining terms which are absent in the corresponding approximations. Being $g(y)=\left[\gamma+\beta\ln(y)\right]y^{2s}$, we get
\begin{equation}
\dot{g}(y)=\left\{\beta+2s\left[\gamma+\beta\ln(y)\right]\right\}y^{2s-1},\quad\ddot{g}(y)=\left\{\left(2s-1\right)\left\{\beta+2s\left[\gamma+\beta\ln(y)\right]\right\}+2s\beta\right\}y^{2s-2},
\end{equation}
so for the zeroth order approximation we have
\begin{equation}\label{zerorodjust}
\frac{1}{\tilde{\eta}}\left|\frac{\left\{\beta+2s\left[\gamma+\beta\ln(y)\right]\right\}}{2\left[\gamma+\beta\ln(y)\right]^\frac{3}{2}}y^{-(s+1)}\right|\ll1,
\end{equation}
and for the first order approximation we get
\begin{equation}
\frac{1}{\tilde{\eta}^{2}}\left|\frac{y^{-2(s+1)}}{\left[\gamma+\beta\ln(y)\right]^{2}}\right|\left|\frac{5}{16}\frac{\left\{\beta+2s\left[\gamma+\beta\ln(y)\right]\right\}^{2}}{\left[\gamma+\beta\ln(y)\right]}-\frac{1}{4}\left\{\left(2s-1\right)\left\{\beta+2s\left[\gamma+\beta\ln(y)\right]\right\}+2s\beta\right\}\right|\ll1\quad ,
\end{equation}
which is clearly verified when $y\rightarrow\infty$ because $-1<s$, for both cases and for the two quantization methods. Here $\gamma$ and $\beta$ are constants which for the zeroth and first orders are found in Eqs. (\ref{eq:wdwWKBia}) and (\ref{eq:PotWKB2}), respectively.

\section{Detailed calculations for the matter part of the WDW equation}

The matter part of the WDW equation, given by Eq.~(\ref{eq:WSWmatterpotential}),
can be rewritten as
\begin{equation}\label{19.1.3}
\frac{\partial^{2}\varphi_{k}(\alpha,z)}{\partial z^{2}}+
\left(\frac{1}{4}z^2-\beta\right)\varphi_{k}(\alpha,z)=0,
\end{equation}
where
\begin{equation}
 \beta=-\frac{1}{2\hbar(\pi G)^{1/2}}\left[\frac{A^{1/2}a_0^3e^{3\alpha}}{6}-
 \frac{E_k}{A^{1/2}a_0^3e^{3\alpha}}\right],
\end{equation}
and
\begin{equation}
 z=\frac{2 a_0^{3/2} e^{3\alpha/2} (\pi G A)^{1/4}}{\hbar^{1/2}}(\phi-\phi_1).
\end{equation}
Eq.~(\ref{19.1.3}) corresponds to Eq.~19.1.3 of page 686 in Ref.~\cite{Abramow}.
Thus, its solutions are the following parabolic cylinder functions:
\begin{equation}\label{sol}
 \varphi_{k}^{(1)}(\alpha,\phi)=W\left(\beta,\, z\right),\quad
 \varphi_{k}^{(2)}(\alpha,\phi)=W\left(\beta,\,- z\right),\quad
 \varphi_{k}^{(3)}(\alpha,\phi)=E\left(\beta,\,z\right),\quad {\rm and}\quad
 \varphi_{k}^{(4)}(\alpha,\phi)=E^*\left(\beta,\,z\right),
\end{equation}
with
\begin{equation}\label{Efunction}
 E(\alpha,\phi)=K^{-1/2} W\left(\beta,\, z\right)+iK^{1/2} W\left(\beta,\, -z\right),
\end{equation}
and 
\begin{equation}\label{K}
 K=\sqrt{1+e^{2\pi\beta}}-e^{\pi\beta}.
\end{equation}

We are interested in the behaviour of these functions for arbitrary values of the field and
very large values of $\alpha$.
It can be noted that $\beta\rightarrow-\infty$ when $\alpha\rightarrow\infty$.
Defining 
\begin{equation}
 X=\sqrt{z^2-4\beta}=\frac{\sqrt{2}\,a_0^{3/2}e^{3\alpha/2}}{\hbar^{1/2}(\pi GA)^{1/4}}
 \sqrt{2\pi GA(\phi-\phi_1)^2+\frac{A}{6}-\frac{E_k}{a_0^6e^{6\alpha}}},
\end{equation}
it can be noted that $X\rightarrow\infty$ when $\alpha\rightarrow\infty$.
Thus, we focus our attention on the case $\beta<0$ and $X^2\gg0$.
The real solutions for this regime are given by
(Eqs.~19.23.10 and 19.23.11 of page 694 in reference \cite{Abramow})
\begin{equation}\label{pcfaprox1}
 W\left(\beta,\,z\right)=\sqrt{2K}e^{v_r}\cos(\pi/4+\theta+v_i),
\end{equation}
and
\begin{equation}
 W\left(\beta,\,-z\right)=\sqrt{2/K}e^{v_r}\sin(\pi/4+\theta+v_i),
\end{equation}
with
\begin{eqnarray}\label{theta}
 \theta&=&\frac{1}{4}zX-\beta\ln\left(\frac{z+X}{2\sqrt{|\beta|}}\right) \nonumber\\
 &=&\frac{a_0^3e^{3\alpha}}{2\hbar \sqrt{\pi\, G A}}
 \left\{\sqrt{\pi\,G A}(\phi-\phi_1)\sqrt{2\pi GA(\phi-\phi_1)^2+\frac{A}{6}-\frac{E_k}{a_0^6e^{6\alpha}}}\right.\nonumber\\
 &+&\left.\left[\frac{A}{6}-\frac{E_k}{a_0^6e^{6\alpha}}\right]
 \ln\left[\frac{\sqrt{\pi\,G A}(\phi-\phi_1)+\sqrt{2\pi GA(\phi-\phi_1)^2+\frac{A}{6}-\frac{E_k}{a_0^6e^{6\alpha}}}}
 {\sqrt{\frac{A}{6}-\frac{E_k}{a_0^6e^{6\alpha}}}}\right]\right\},
\end{eqnarray}
and $v_i$ and $v_r$ are defined by series 
that go as $v_r\approx-1/2\,\ln X$ and $v_i\approx0$ for very large values of $X$
(see Eq.~19.23.4 of Ref.~\cite{Abramow}).
Moreover, using Eqs.~(\ref{K}) and (\ref{theta}), 
it can be seen that $\theta$ diverges as $e^{3\alpha}$ and $K\rightarrow1$ for $\alpha\rightarrow\infty$.
For large values of $\alpha$, we can approximate Eq.~(\ref{pcfaprox1}) as
\begin{equation}\label{pcfalpha}
 W\left(\beta,\,z\right)\approx\sqrt{2}\,X^{-1/2}\cos\left[\frac{\pi}{4}+\frac{1}{4}zX-\beta\ln\left(\frac{z+X}{2\sqrt{|\beta|}}\right)\right].
\end{equation}
Thus, we have
\begin{equation}\label{limit1}
 W\left(\beta,\,z\right)\sim e^{-3\alpha/4}\cos(e^{3\alpha})\rightarrow0,
\end{equation}
and
\begin{equation}\label{limit2}
 W\left(\beta,\,-z\right)\sim e^{-3\alpha/4}\sin(e^{3\alpha})\rightarrow0,
\end{equation}
for $\alpha\rightarrow\infty$. On the other hand, taking into account
Eqs.~(\ref{limit1}) and (\ref{limit2}) into Eq.~(\ref{Efunction}),
one can conclude that the imaginary solutions also vanish in the asymptotic limit.

\section{Validity of the Born--Oppenheimer approximation}

Consider the WDW equation (\ref{eq:wdweqBornOp}) for the phantom case,

\begin{equation}
\frac{\hbar^{2}}{2}\frac{\kappa^{2}}{6}\left(C_{k}\frac{\partial^{2}\varphi_{k}}{\partial\alpha^{2}}+\varphi_{k}\frac{\partial^{2}C_{k}}{\partial\alpha^{2}}+2\frac{\partial C_{k}}{\partial\alpha}\frac{\partial\varphi_{k}}{\partial\alpha}\right)+\frac{\hbar^{2}}{2}C_{k}\frac{\partial^{2}\varphi_{k}}{\partial\phi^{2}}+a_{0}^{6}e^{6\alpha}V(\phi)C_{k}\varphi_{k}=0.
\end{equation}
The BO approximation assumes that the first and third terms of this equation can be neglected as
compared with the second term. That is, it considers that the variations with respect
to geometry of the geometric part, $C_{k}(\alpha)$, are much more important than those
of the matter part, $\varphi_{k}(\alpha,\phi)$, and that, therefore, there is a negligible backreaction
of the matter part on the gravitational part. 

The validity of this approximation can be explored using expansions with respect to $\kappa$ (see Ref. \cite{Bouhmadi-Lopez:2009pu} and references cited there).
For the geometric part, from  Eq. (\ref{eq:wdwgeometric}), we get
\begin{equation}
C_{k}\sim\mathcal{O}(\kappa^0),\quad\dot{C_{k}}\sim\mathcal{O}(\kappa^{-1}),
\quad\ddot{C_{k}}\sim\mathcal{O}(\kappa^{-2}),
\end{equation}
with $\dot{}\equiv\partial/\partial\alpha$.
The matter part is given in terms of parabolic cylinder functions. In order to get its expansion in terms of $\kappa$ and
$A$ in a simple way, we use the approximation for large values of the scale factor given by Eq.~(\ref{pcfalpha}).
We should also keep track of the dependence on $\alpha$.
Noting that
\begin{equation}
 \beta\sim\theta\sim \mathcal{O}(\kappa^{-1}A^{1/2}a_0^3)\,e^{3\alpha},\quad 
 z\sim X\sim\mathcal{O}(\kappa^{-1/2}A^{1/4}a_0^{3/2})e^{3\alpha/2},
\end{equation}
we get
\begin{equation}
 \varphi_k(\alpha,\phi)\sim  E(\alpha,\phi)=\sqrt{2}\,\exp\left[h(\alpha,\phi)\right]
\end{equation}
where 
\begin{equation}
 h(\alpha)\sim \mathcal{O}\left(1\right)\,\alpha+
 \mathcal{O}\left(\kappa^{-1}A^{1/2}a_0^{3}\right)e^{3\alpha}
\end{equation}
and the quantity $\kappa^{-1/2}A^{1/4}a_0^{3/2}$ is dimensionless.
Thus, we have
\begin{equation}
\frac{C_{k}\ddot\varphi_{k}}{\ddot{C_{k}}\varphi_{k}}\sim 
\mathcal{O}(\kappa^{2})\left[\dot h^2+\ddot h\right]
\sim \mathcal{O}(\kappa^2)+\mathcal{O}(\kappa\, A^{1/2}\,a_0^3)\,e^{3\alpha}+
\mathcal{O}(A\,a_0^6)\,e^{6\alpha},
\end{equation}
and
\begin{equation}
\frac{\dot C_{k}\dot\varphi_{k}}{\ddot{C_{k}}\varphi_{k}}\sim \mathcal{O}(\kappa)\, \dot h
\sim \mathcal{O}(\kappa^1)+\mathcal{O}(A^{1/2}a_0^3)\,e^{3\alpha}.
\end{equation}
Thus, the B.O. approximation is fulfilled as long as $Aa_0^6e^{6\alpha} \ll\kappa^2$; i.e. this approach is valid for sufficiently small value of $A$, i.e. the fluid (\ref{eq:eostate}) is close enough to a cosmological constant and for large enough values of $a$ but not infinite value of $a$, that is within a semiclassical regime.


\begin{thebibliography}{10}

%
  \bibitem{Bouhmadi-Lopez:2014cca}
  M.~Bouhmadi-L\'{o}pez, A.~Errahmani, P.~Mart\'{\i}n-Moruno, T.~Ouali and Y.~Tavakoli,
  %``The little sibling of the big rip singularity,''
  Int.\ J.\ Mod.\ Phys.\ D {\bf 24} (2015) 1550078
  [arXiv:1407.2446 [gr-qc]].

%\cite{Hawking:1969sw}
\bibitem{Penrose}
  S.~W.~Hawking and R.~Penrose,
  %``The Singularities of gravitational collapse and cosmology,''
  Proc.\ Roy.\ Soc.\ Lond.\ A {\bf 314} (1970) 529.
  %%CITATION = PRSLA,A314,529;%%
  %531 citations counted in INSPIRE as of 22 sept. 2015

  \bibitem{Penrose and Hawk}%3%
  S.~W.~Hawking and R.~Penrose, The Nature of Space and Time, Princeton University Press (1996).
  
%\cite{Weinberg:2000yb}
\bibitem{Weinberg:2000yb}
  S.~Weinberg,
  %``The Cosmological constant problems,''
  astro-ph/0005265.
  %%CITATION = ASTRO-PH/0005265;%%
  %304 citations counted in INSPIRE as of 22 sept. 2015

  
    \bibitem{Peebles:2002gy}%6%
  P.~J.~E.~Peebles and B.~Ratra,
  %``The Cosmological constant and dark energy,''
  Rev.\ Mod.\ Phys.\  {\bf 75} (2003) 559
  [astro-ph/0207347].
  
  \bibitem{Padmanabhan:2002ji}%7%
  T.~Padmanabhan,
  %``Cosmological constant: The Weight of the vacuum,''
  Phys.\ Rept.\  {\bf 380} (2003) 235
  [hep-th/0212290].  
  
  \bibitem{Sahnidmde}%4%
  V.~Sahni,
  Lect.\ Notes Phys.\  {\bf 653} (2004) 141
  [astro-ph/0403324].
  
  \bibitem{Copeland:2006wr}%8%
  E.~J.~Copeland, M.~Sami and S.~Tsujikawa,
  %``Dynamics of dark energy,''
  Int.\ J.\ Mod.\ Phys.\ D {\bf 15} (2006) 1753
  [hep-th/0603057].

  \bibitem{Frieman:2008sn}%9%
  J.~Frieman, M.~Turner and D.~Huterer,
  %``Dark Energy and the Accelerating Universe,''
  Ann.\ Rev.\ Astron.\ Astrophys.\  {\bf 46} (2008) 385
  [arXiv:0803.0982 [astro-ph]]

%\cite{Ade:2015xua}
\bibitem{Ade:2015xua}
  P.~A.~R.~Ade {\it et al.} [Planck Collaboration],
  %``Planck 2015 results. XIII. Cosmological parameters,''
  arXiv:1502.01589 [astro-ph.CO].
  %%CITATION = ARXIV:1502.01589;%%
  %657 citations counted in INSPIRE as of 22 sept. 2015

%\cite{Sahni:1999gb}
\bibitem{Sahni:1999gb} 
  V.~Sahni and A.~A.~Starobinsky,
  %``The Case for a positive cosmological Lambda term,''
  Int.\ J.\ Mod.\ Phys.\ D {\bf 9}, 373 (2000)
  [astro-ph/9904398].
  %%CITATION = ASTRO-PH/9904398;%%
  %1475 citations counted in INSPIRE as of 28 juil. 2015
   
   
  \bibitem{Kamenshchik:2001cp}
  A.~Y.~Kamenshchik, U.~Moschella and V.~Pasquier,
  %``An Alternative to quintessence,''
  Phys.\ Lett.\ B {\bf 511} (2001) 265
  [gr-qc/0103004].

  \bibitem{Bento:2003dj}
  M.~C.~Bento, O.~Bertolami and A.~A.~Sen,
  %``Generalized Chaplygin gas model: Dark energy - dark matter unification and CMBR constraints,''
  Gen.\ Rel.\ Grav.\  {\bf 35} (2003) 2063
  [gr-qc/0305086].
  
  \bibitem{Bouhmadi-Lopez:2004me}
  M.~Bouhmadi-L\'{o}pez and J.~A.~Jimenez Madrid,
  %``Escaping the big rip?,''
  JCAP {\bf 0505} (2005) 005
  [astro-ph/0404540].
 
 \bibitem{Nojiri:2005pu}
  S.~Nojiri and S.~D.~Odintsov,
  %``Unifying phantom inflation with late-time acceleration: Scalar phantom-non-phantom transition model and generalized holographic dark energy,''
  Gen.\ Rel.\ Grav.\  {\bf 38} (2006) 1285
  [hep-th/0506212].
  
  \bibitem{Costa:2013sva}
  A.~A.~Costa, X.~D.~Xu, B.~Wang, E.~G.~M.~Ferreira and E.~Abdalla,
  %``Testing the Interaction between Dark Energy and Dark Matter with Planck Data,''
  Phys.\ Rev.\ D {\bf 89} (2014) 10,  103531
  [arXiv:1311.7380 [astro-ph.CO]].
 
  \bibitem{Salvatelli:2014zta}
  V.~Salvatelli, N.~Said, M.~Bruni, A.~Melchiorri and D.~Wands,
  %``Indications of a late-time interaction in the dark sector,''
  Phys.\ Rev.\ Lett.\  {\bf 113} (2014) 18,  181301
  [arXiv:1406.7297 [astro-ph.CO]]   
      
   \bibitem{StarobBigrip}
   A.~ A.~Starobinsky, Grav. Cosmol.{\bf 6} (2000) 157
     
   \bibitem{Caldwellbigrip}
    R.~R.~Caldwell, M. Kamionkowski and N. N. Weinberg,
    Phys.\ Rev.\ Lett.\  {\bf 91}  (2003) 071301.
   
   \bibitem{Caldwellbigrip2}
   R.~R.~Caldwell, Phys.\ Lett.\ B {\bf 545} (2002) 23
   
   \bibitem{Caldwell:2003vq}
  R.~R.~Caldwell, M.~Kamionkowski and N.~N.~Weinberg,
  %``Phantom energy and cosmic doomsday,''
  Phys.\ Rev.\ Lett.\  {\bf 91} (2003) 071301
  [astro-ph/0302506].
   
  \bibitem{Carrollbigrip}
  S.~M.~Carroll, M.~Hoffman and M.~Trodden,  Phys.\ Rev.\ 
  D {\bf 68} (2003) 023509.

  \bibitem{Chimentobigrip}
  L.~P.~Chimento and R.~Lazkoz, Phys.\ Rev.\ Lett.\  {\bf 91}
  (2003) 211301.

  \bibitem{Dabrbigrip}
  M.~P.~D\c{a}browski, T.~Stachowiak and M.~Szydlowski,
  Phys.\ Rev.\ D {\bf 68} (2003) 103519.


\bibitem{Gonzalezbigrip}
  P.~F.~Gonz\'{a}lez-D\'{i}az,
  %``K-essential phantom energy: Doomsday around the corner?,''
  Phys.\ Lett.\ B {\bf 586} (2004) 1
  [astro-ph/0312579].
  %%CITATION = ASTRO-PH/0312579;%%
  %206 citations counted in INSPIRE as of 22 sept. 2015


  \bibitem{Gonzalezbigrip2}
  P.~F.~Gonz\'{a}lez-D\'{i}az,  Phys.\ Rev.\ D {\bf 69} (2004) 063522

\bibitem{Barrowbigbrake}
  J.~D.~Barrow, G.~J.~Galloway, and F.~J.~Tipler, Mon.\ Not.\ R.~astr.\ Soc. 223, (1986) 835-844.
  
    \bibitem{Barrowsudden}
  J.~D.~Barrow, Class. Quant.\ Grav.\ {\bf 21} (2004) L79.

  
%\cite{Gorini:2003wa}
\bibitem{Gorinibigbrake}
  V.~Gorini, A.~Y.~Kamenshchik, U.~Moschella and V.~Pasquier,
  %``Tachyons, scalar fields and cosmology,''
  Phys.\ Rev.\ D {\bf 69} (2004) 123512
  [hep-th/0311111].
  %%CITATION = HEP-TH/0311111;%%
  %210 citations counted in INSPIRE as of 22 sept. 2015
           

  \bibitem{Nojirisudden}                 
  S.~Nojiri, S.~D.~Odintsov and S.~Tsujikawa, Phys.\ Rev.\
  D {\bf 71} (2005) 063004.  
                       
  
    \bibitem{Mariambigfreeze}
  M.~Bouhmadi-L\'{o}pez, P.~F. Gonz\'{a}lez-D\'{i}az and P.~Mart\'{\i}n-Moruno,
  Int.\ J.\ Mod.\ Phys.\ D {\bf 17} (2008) 2269.

  \bibitem{Mariambigfreeze2}
  M.~Bouhmadi-L\'{o}pez, P.~F.~Gonz\'{a}lez-D\'{i}az and P.~Mart\'{\i}n-Moruno,
  Phys.\ Lett.\ B {\bf 659} (2008) 1.


  \bibitem{Nojiribigfreeze}                                                                                                                  
  S.~Nojiri and S.~D.~Odintsov, Phys.\ Rev.\ D {\bf 70} (2004)
  103522.

  \bibitem{Nojiribigfreeze2}
  S.~Nojiri and S.~D.~Odintsov, Phys.\ Rev.\ D {\bf 72} (2005)
  023003.

  \bibitem{Nojiritypeiv}
  S.~Nojiri and S.~D.~Odintsov, Phys.\ Rev.\ D {\bf 78} (2008)
  046006.

  \bibitem{Bambatypeiv}
  K.~Bamba, S.~Nojiri and S.~D.~Odintsov, JCAP 0810
  (2008) 045.
  
    \bibitem{Nojiri:2015fia}
  S.~Nojiri, S.~D.~Odintsov, V.~K.~Oikonomou and E.~N.~Saridakis,
  %``Singular cosmological evolution using canonical and phantom scalar fields,''
  arXiv:1503.08443 [gr-qc].
                                                                                            
  \bibitem{Ruzmaikina1970}
  T. Ruzmaikina and A. A. Ruzmaikin, Sov. Phys. JETP {\bf 30}, 372 (1970).

  \bibitem{Stefancic:2004kb}
  H.~\v{S}tefan\v{c}i\'{c},
  %``Expansion around the vacuum equation of state - Sudden future singularities and asymptotic behavior,''
  Phys.\ Rev.\ D {\bf 71} (2005) 084024.
  %%CITATION = ASTRO-PH/0411630;%%
  %119 citations counted in INSPIRE as of 17 juin 2015

  \bibitem{BouhmadiLopez:2005gk}
  M.~Bouhmadi-L\'{o}pez,
  %``Phantom-like behaviour in dilatonic brane-world scenario with induced gravity,''
  Nucl.\ Phys.\ B {\bf 797} (2008) 78.
  %%CITATION = ASTRO-PH/0512124;%%
  %43 citations counted in INSPIRE as of 17 Jun 2015

  \bibitem{Frampton:2011sp}
  P.~H.~Frampton, K.~J.~Ludwick and R.~J.~Scherrer,
  %``The Little Rip,''
  Phys.\ Rev.\ D {\bf 84} (2011) 063003.
  %%CITATION = ARXIV:1106.4996;%%
  %102 citations counted in INSPIRE as of 16 juin 2015

  \bibitem{Brevik:2011mm}
  I.~Brevik, E.~Elizalde, S.~Nojiri and S.~D.~Odintsov,
  %``Viscous Little Rip Cosmology,''
  Phys.\ Rev.\ D {\bf 84} (2011) 103508.
  %%CITATION = ARXIV:1107.4642;%%
  %78 citations counted in INSPIRE as of 16 juin 2015

  \bibitem{Bouhmadi-Lopez:2013nma}
  M.~Bouhmadi-L\'{o}pez, P.~Chen and Y.~-W.~Liu,
  %``Tradeoff between smoother and sooner 'little rip',''
  Eur.\ Phys.\ J.\ C {\bf 73} (2013) 9,  2546.
  %%CITATION = ARXIV:1302.6249;%%
  %5 citations counted in INSPIRE as of 17 Jun 2015
  
     \bibitem{Dabrowski:2006dd}
  M.~P.~D\c{a}browski, C.~Kiefer and B.~Sandh\"{o}fer,
  %``Quantum phantom cosmology,''
  Phys.\ Rev.\ D {\bf 74} (2006) 044022
  [hep-th/0605229].
  
     \bibitem{Kamenshchik:2007zj}
  A.~Kamenshchik, C.~Kiefer and B.~Sandh\"{o}fer,
  %``Quantum cosmology with big-brake singularity,''
  Phys.\ Rev.\ D {\bf 76} (2007) 064032
  [arXiv:0705.1688 [gr-qc]].
                                                                                                                                                                                                                                                                                                                                                                                                                                                                                                                                                                                                                                                                                                
   \bibitem{Bouhmadi-Lopez:2009pu}
  M.~Bouhmadi-L\'{o}pez, C.~Kiefer, B.~Sandh\"{o}fer and P.~V.~Moniz,
  %``On the quantum fate of singularities in a DE dominated universe,''
  Phys.\ Rev.\ D {\bf 79} (2009) 124035
  [arXiv:0905.2421 [gr-qc]].
    
  \bibitem{Bouhmadi-Lopez:2013tua}
  M.~Bouhmadi-L\'{o}pez, C.~Kiefer and M.~Kr\"{a}mer,
  %``Resolution of type IV singularities in quantum cosmology,''
  Phys.\ Rev.\ D {\bf 89} (2014) 6,  064016
  [arXiv:1312.5976 [gr-qc]].
  
%\cite{Albarran:2015tga}
\bibitem{Albarran:2015tga}
  I.~Albarran and M.~Bouhmadi-L\'{o}pez,
  %``Quantisation of the holographic Ricci dark energy model,''
  JCAP {\bf 1508} (2015) 08,  051
  [arXiv:1505.01353 [gr-qc]].
  %%CITATION = ARXIV:1505.01353;%%

     
  \bibitem{deHaro:2012xj}
  J.~de Haro,
  %``Does loop quantum cosmology replace the big rip singularity by a non-singular bounce?,''
  JCAP {\bf 1211} (2012) 037
  [arXiv:1207.3621 [gr-qc]].
  
    \bibitem{Bamba:2012ka}
  K.~Bamba, J.~de Haro and S.~D.~Odintsov,
  %``Future Singularities and Teleparallelism in Loop Quantum Cosmology,''
  JCAP {\bf 1302} (2013) 008
  [arXiv:1211.2968 [gr-qc]].
   
  \bibitem{Barrow:2011ub}
  J.~D.~Barrow, A.~B.~Batista, J.~C.~Fabris, M.~J.~S.~Houndjo and G.~Dito,
  %``Sudden singularities survive massive quantum particle production,''
  Phys.\ Rev.\ D {\bf 84} (2011) 123518
  [arXiv:1110.1321 [gr-qc]].
  
  \bibitem{Barrow:2015sga}
  J.~D.~Barrow and A.~A.~H.~Graham,
  %``New Singularities in Unexpected Places,''
  arXiv:1505.04003 [gr-qc].
  
  \bibitem{Elizalde:2004mq}
  E.~Elizalde, S.~Nojiri and S.~D.~Odintsov,
  %``Late-time cosmology in (phantom) scalar-tensor theory: Dark energy and the cosmic speed-up,''
  Phys.\ Rev.\ D {\bf 70} (2004) 043539
  [hep-th/0405034].
  
  \bibitem{KieferQG}
  C.~Kiefer, \textit{Quantum Gravity}. Second edition (Oxford University Press, Oxford, 2007). 
  
  \bibitem{PMonizQC}
  P.~Moniz, \textit{Quantum Cosmology - The Supersymmetric Perspective - Vol.1: Fundamentals}. Lect. Notes Phys. 803 (Springer, Berlin Heidelberg   2010). 
  
  \bibitem{FuturePhscs}
  J.~Hartle, \textit{The state of the universe}, pp. 615\textendash 620, C. Page, \textit{Quantum Cosmology}, pp. 621\textendash 648, A. Vilenkin,  \textit{Quantum Cosmology and eternal inflation}, pp. 649\textendash 666, in   \textit{The    future of Theoretical Physics and Cosmology. Celebrating    Stephen Hawking's 60th Birthday}, edited by G.w. Gibbons, E. P. S. Shellard and S. J. Rankin (Cambridge University Press, Cambridge, 2003).
  
    %\cite{BouhmadiLopez:2004mp}
\bibitem{BouhmadiLopez:2004mp} 
  M.~Bouhmadi-L\'{o}pez and P.~Vargas Moniz,
  %``FRW quantum cosmology with a generalized Chaplygin gas,''
  Phys.\ Rev.\ D {\bf 71}, 063521 (2005)
  [gr-qc/0404111].
  %%CITATION = GR-QC/0404111;%%
  %47 citations counted in INSPIRE as of 29 juil. 2015
    
  %\cite{BouhmadiLopez:2006pf}
\bibitem{BouhmadiLopez:2006pf}
  M.~Bouhmadi-L\'{o}pez and P.~Vargas Moniz,
  %``Quantisation of Parameters and the String Landscape Problem,''
  JCAP {\bf 0705} (2007) 005
  [hep-th/0612149].
  %%CITATION = HEP-TH/0612149;%%
  %10 citations counted in INSPIRE as of 29 Jul 2015
  
  
    \bibitem{Mathew}
  J.~Mathews and R. L.~Walker, \textit{Mathematical methods of Physics} (California Institute of Techonolgy, 1969).
  
%\cite{Barvinsky:1993jf}
\bibitem{Barvinsky:1993jf}
  A.~O.~Barvinsky,
  %``Unitarity approach to quantum cosmology,''
  Phys.\ Rept.\  {\bf 230} (1993) 237.
  %%CITATION = PRPLC,230,237;%%
  %90 citations counted in INSPIRE as of 22 sept. 2015

%\cite{Kamenshchik:2012ij}
\bibitem{Kamenshchik:2012ij}
  A.~Y.~Kamenshchik and S.~Manti,
  %``Classical and quantum Big Brake cosmology for scalar field and tachyonic models,''
  Phys.\ Rev.\ D {\bf 85} (2012) 123518
  [arXiv:1202.0174 [gr-qc]].
  %%CITATION = ARXIV:1202.0174;%%
  %9 citations counted in INSPIRE as of 22 sept. 2015

%\cite{Kamenshchik:2013naa}
\bibitem{Kamenshchik:2013naa}
  A.~Y.~Kamenshchik,
  %``Quantum cosmology and late-time singularities,''
  Class.\ Quant.\ Grav.\  {\bf 30} (2013) 173001
  [arXiv:1307.5623 [gr-qc]].
  %%CITATION = ARXIV:1307.5623;%%
  %8 citations counted in INSPIRE as of 22 sept. 2015





%\cite{Barvinsky:2013aya}
\bibitem{Barvinsky:2013aya}
  A.~O.~Barvinsky and A.~Y.~Kamenshchik,
  %``Selection rules for the Wheeler-DeWitt equation in quantum cosmology,''
  Phys.\ Rev.\ D {\bf 89} (2014) 4,  043526
  [arXiv:1312.3147 [gr-qc]].
  %%CITATION = ARXIV:1312.3147;%%
  %5 citations counted in INSPIRE as of 22 sept. 2015


%\cite{Kuo:1993if}
\bibitem{Kuo:1993if}
  C.~I.~Kuo and L.~H.~Ford,
  %``Semiclassical gravity theory and quantum fluctuations,''
  Phys.\ Rev.\ D {\bf 47} (1993) 4510
  [gr-qc/9304008].
  %%CITATION = GR-QC/9304008;%%
  %113 citations counted in INSPIRE as of 22 sept. 2015

  
  %\cite{Martin-Moruno:2013wfa}
\bibitem{Martin-Moruno:2013wfa}
  P.~Mart\'{\i}n-Moruno and M.~Visser,
  %``Semiclassical energy conditions for quantum vacuum states,''
  JHEP {\bf 1309} (2013) 050
  [arXiv:1306.2076 [gr-qc]]
  
  %\cite{Martin-Moruno:2013sfa}
\bibitem{Martin-Moruno:2013sfa}
  P.~Mart\'{\i}n-Moruno and M.~Visser,
  %``Classical and quantum flux energy conditions for quantum vacuum states,''
  Phys.\ Rev.\ D {\bf 88} (2013) 6,  061701
  [arXiv:1305.1993 [gr-qc]].
  
\bibitem{Bousso:2015wca}
  R.~Bousso, Z.~Fisher, J.~Koeller, S.~Leichenauer and A.~C.~Wall,
  %``Proof of the Quantum Null Energy Condition,''
  arXiv:1509.02542 [hep-th].
  %%CITATION = ARXIV:1509.02542;%%  
  
  \bibitem{Abramow}
  M.~Abramowitz and I.~Stegun, \textit{Handbook on Mathematical Functions} (Dover, 1980).


 
 \end{thebibliography}
\end{document}